\documentclass[aps,pra,groupedaddress,superscriptaddress,amsmath,floatfix,nofootinbib,reprint]{revtex4-1}
\usepackage{graphicx}
\usepackage{bm}
\usepackage[english]{babel}
\usepackage{xspace}
\usepackage{multirow} 
\usepackage{array}
\usepackage[tight,nice]{units}

\newcommand{\ion}[3]{{#1}\ensuremath{^{{#2}{#3}}}\xspace}
\newcommand{\erbium}{\ion{Er}{3}{+}}%{Er$^{3+}$}%{\ion{Er}{3}{+}{}}

\newcommand{\enlev}[4]{\ensuremath{^{#1}#2_{\nicefrac{#3}{#4}}}}%Command allows easy typing of Russell-Saunders notation. Typed as it is said, ie, '{4}{I}{13}(over){2}' for the first excited state of erbium.
\newcommand{\etal}{~\emph{et~al.}\xspace}
\newcommand{\dgr}{\ensuremath{^\circ}}
\newcommand{\dgra}{\ensuremath{^\circ}\xspace}

\usepackage{miller}%Typing Miller indices. Usage: \hkl{xyz} or \hkl[xyz] or \hkl<xyz> or \hkl(xyz) where x,y,z are pos or neg

\begin{document}

\title{Theoretical and experimental evaluation of multilayer porous silicon structures for enhanced erbium up-conversion luminescence}

\author{Craig M. Johnson}
\email[Corresponding author: ]{craig.johnson@unsw.edu.au}
\affiliation{School of Photovoltaic and Renewable Energy Engineering, University of New South Wales \\
Sydney 2052 NSW Australia}
\affiliation{Optoelectronics Group, School of Physics, University of New South Wales \\
Sydney 2052 NSW Australia}
\author{Peter J. Reece}
\affiliation{Optoelectronics Group, School of Physics, University of New South Wales \\
Sydney 2052 NSW Australia}
%\author{H. Hoe Tan}
%\affiliation{Department of Electronic Materials Engineering, Research School of Physics and Engineering, Australian National University \\
%Canberra 0200 ACT Australia}
\author{Gavin J. Conibeer}
\affiliation{School of Photovoltaic and Renewable Energy Engineering, University of New South Wales \\
Sydney 2052 NSW Australia}

\date{\today}

\begin{abstract}	
The enhancement of \erbium-based up-conversion for photovoltaics in multilayer porous silicon photonic structures is considered theoretically and experimentally. Transfer matrix simulations are used to assess the increased photonic density of states that results from the slowing of energy propagation at the short-wavelength edge of one-dimensional photonic band gaps. An indirect calculation of \erbium absorption enhancement within slow-light modes is then used to illustrate an increase in absorption over the bulk value: the effective absorption coefficient is shown to increase by more than 22\% over a broad spectral region and by more than 400\% over a narrow region. Erbium-doped porous silicon photonic crystals are fabricated with the photonic band edge coincident with the \enlev{4}{I}{15}{2}$\rightarrow$\enlev{4}{I}{13}{2} \erbium transition. Challenges in fabrication %and the results of compositional analysis 
are discussed. An angular-dependent photoluminescence measurement demonstrates emission intensity that varies non-monotonically with the position of the photonic band edge. A maximum of 26.6$\times$ enhancement of \erbium emission intensity is observed for the 550-nm transition, with lower enhancement factors seen for longer wavelengths. 
\end{abstract}


\maketitle 
\section{Introduction}
The up-conversion (UC) of the substantial sub-band-gap component of the solar spectrum to shorter wavelengths may provide the basis for a commercial photovoltaic (PV) device that overcomes the theoretical limiting efficiency of conventional single-junction solar cells~\cite{Trupke2002,Conibeer2008a,Ende2009,Wild2011,Atre2011}. A simple UC-PV device would comprise a standard single-junction solar cell with an additional UC layer to modify the incident spectrum~\cite{Strumpel2007}. Lanthanide-doped phosphors (e.g., NaYF$_{4}$:\erbium) have been the subject of intensive study for UC layers coupled to both crystalline and amorphous Si solar cells~\cite{Shalav2007,Ivanova2009,Wild2011}. Phosphors are typically low-phonon-energy hosts with sufficient electrostatic site asymmetry at the location of the rare earth ion to relax parity selection rules~\cite{Auzel2004}. This allows for long excited state lifetimes and strong luminescence in the visible and near-infrared spectral regions when excited with near-infrared light. However, substantial efficiency gains at reasonable concentration ratios---particularly in c-Si cells---have yet to be reported for phosphor-based devices, due in part to the weak and narrow-band absorption of rare earth ions. Improvement of the optical properties of common UC materials is therefore the current challenge for UC-PV device design.

Most studies of UC optimization to date have focused on the synthesis and characterization of \erbium-doped UC phosphors for a wide variety of applications~\cite{Ende2009,Vennerberg2011,Renero2011,Liang2011, Sarakovskis2011,Shan2011,Tian2011}. Other important investigations have considered co-doping of \erbium with other rare earth elements~\cite{Wild2010,Wang2011} or sensitizing of \erbium absorption with local broad-band absorbers, e.g., semiconductor nanocrystals~\cite{Lucarz2003,Jensen2006a, Das2007a,Rogach2007,Izeddin2008,Fischer2008,Guzman2009,Heng2009}. Recently, the enhancement of UC efficiency by the plasmonic modification of local electromagnetic fields has attracted considerable attention~\cite{Aisaka2008,Deng2011,Vennerberg2011,Atre2012,Fischer2012}. These approaches are promising: for example, Deng\etal reported a nine-fold emission enhancement for NIR-to-visible UC in NaYF$_4$:\erbium nanoparticles coupled to noble metal particles~\cite{Deng2011}.

Building on such studies of \emph{photonic} enhancement of UC efficiency, we report here on the investigation of Si-based photonic crystals (PCs) as an alternative means of modifying electromagnetic fields in the vicinity of UC species. One-dimensional photonic crystals are characterized by a band of optical modes that are unable to propagate within the crystal---a ``photonic band gap'' (PBG)---due to the discrete translational invariance of the structure as defined by its refractive index profile~\cite{Joannopoulos2008}. Near the edges of the PBG, in the transitional region between high transmission and total reflection, the velocity of energy propagation through the crystal slows down significantly~\cite{Krauss2007}. The conservation of energy flux dictates that the energy density in these slow modes must increase in proportion to the relative decrease of the propagation velocity~\cite{Boyd2011}. 

While we have previously shown~\cite{Johnson2011} that high field intensity is apparent at the band edge of one-dimensional photonic structures, the impact this would have on \erbium absorption was implied but not directly demonstrated. In this paper we theoretically assess the potential for enhanced coupling of incident light to \erbium ions situated within these PCs when the slow light band-edge mode is designed to coincide with the intrinsic absorption of \erbium. This is followed by a preliminary experimental study of band-edge UC luminescence in an \erbium-doped porous silicon (PSi:\erbium) PC.

\section{Optical characteristics of multilayer dielectric stacks}
It was already clear to Rayleigh~\cite{Rayleigh1887,Rayleigh1917} 130 years ago that the wave properties of light make perfect reflectance possible in stratified dielectric media over all spectral ranges due to interference phenomena. In a single thin film, the phase shifts imparted on propagating waves at dielectric interfaces may result in constructive and destructive interference; for repeating layers of alternating thickness and refractive index, many interferences may coherently interact to produce high reflectivity at certain points in the spectrum. The condition by which complete reflectivity is possible is the Bragg condition (for first order diffraction):
\begin{equation}
\lambda = 2nd\sin\theta.
\end{equation}

For a single thin film with thickness $d$ and refractive index $n$, this condition produces a reflectivity peak at $\lambda$ but not perfect reflectivity. When a periodicity is enforced by alternating between two layers of thickness $d_i$ and refractive index $n_i$ such that
\begin{equation}
n_1d_1=n_2d_2=\lambda/4,
\label{eq: 1/4WL}
\end{equation}
peak reflectivity increases rapidly with the number of bilayers $N$ and is given by~\cite{Yeh1988}
\begin{equation}
R\simeq\left(\frac{1-(n_s/n_a)(n_2/n_1)^{2N}}{1+(n_s/n_a)(n_2/n_1)^{2N}}\right)^2
\label{eq: DBRRef}
\end{equation}
where it is assumed that $n_2>n_1$ and the substrate refractive index $n_s$ is similar to that of the layers; $n_a$ is the refractive index of the ambient. While this reflectivity is achieved for the design wavelength $\lambda_c$, for a sufficient refractive index contrast $n_2/n_1$ a wide band of perfect reflectivity centered at $\lambda_c$ may be observed. The spectral width of the PBG is given by 
\begin{equation}
\frac{\Delta\lambda}{\lambda_c}=\frac{4}{\pi}\sin^{-1}{\frac{n_2-n_1}{n_2+n_1}}.
\label{eq: DBRwidth}
\end{equation}

These multilayer dielectric reflectors are often called \emph{distributed Bragg reflectors} (DBRs) and have found wide application; the optical properties of DBRs may be calculated by a transfer matrix method~\cite{Yeh1977,Katsidis2002}. Bragg reflectors are fabricated for commercial and research purposes from a range of materials, often by time-consuming epitaxial (or other vacuum) techniques~\cite{Seurin2012}. DBRs based on sputtered SiC layers have been suggested as selective reflectors for photon management in UC-PV devices~\cite{Herter2011}. 

Porous silicon (PSi) presents an alternative to such processes, affording the ability to rapidly fabricate high-quality, high-contrast multilayer structures at room temperature and atmospheric pressure~\cite{Lopez2000a,Reece2004,Reece2004b,Ilyas2007}. In a PSi DBR, the refractive index of the material is proportional to the porosity, which may be periodically modulated by adjusting the etching current during fabrication. This results in a planar multilayer structure with refractive index variation in one dimension~\cite{Setzu1998}. Such structures may be doped with \erbium or other ionic species by electroplating of the large internal surface~\cite{Kimura1994}; this method has been shown to result in \erbium luminescence after annealing in oxygen~\cite{Lopez2000a,Squire2000,Lopez2001}. 

The optical properties of PSi:\erbium DBRs may be simulated using effective medium approximations~\cite{Khardani2007,Niklasson1981} as inputs to the transfer matrix method~\cite{Yeh1977,Katsidis2002} for a multilayer structure.

\section{Modeling porous silicon multilayer optical structures}
In order to study the optical absorption properties of PSi multilayer structures, a series of seven DBRs was simulated by the transfer matrix method. The DBRs were designed with an increasing number of high- and low-porosity bilayers with porosities fixed at $P_H$=0.7 and $P_L$=0.5, respectively, to correspond to achievable experimental conditions. For each DBR, the center wavelength of the stop band $\lambda_c$ was selected by trial-and-error so that the short-wavelength transmission peak (i.e., the photonic band edge) was located at 1550$\pm$0.5\,nm; while this is not the maximum \erbium absorption wavelength, it is the wavelength of the laser used in the subsequent experiment and is within the region of strong \erbium absorption. A choice of $\lambda_c$ fixes the layer thicknesses $d_H$ and $d_L$ by the quarter-wavelength condition in Eq.~\ref{eq: 1/4WL}. 

Table~\ref{tab: MLDP} shows the design parameters for the simulated DBRs: the number of bilayers $N$ ranges from 10 to 100, so that the structure thickness increases by approximately ten times. To investigate the optical properties of an analogous material \emph{without} a photonic stop band, an eighth structure was simulated with eighty ``bilayers'' of a single porosity; it is therefore designated $H^{80}$ in Table~\ref{tab: MLDP}. The porosity of this structure was taken to be 0.6---the average of the porosities used in the DBRs---and the layer thicknesses were designed so that negligible reflectivity is obtained near 1550\,nm. The total thickness of $H^{80}$ is approximately equal to that of (\emph{HL})$^{40}$.

\begin{table*}
\centering
\caption{Design parameters of PSi DBRs with band edge at 1550\,nm (normal incidence)}
\label{tab: MLDP}
\begin{tabular}{|*{6}{c|}c|}
\cline{1-7}
\multirow{2}{*}{ \bfseries Structure }&\multirow{2}{*}{ $\bm{P_H}$ }&\multirow{2}{*}{ $\bm{P_L}$ }&\multirow{2}{*}{ $\bm{\lambda_c}$\bfseries (nm) }&\multirow{2}{*}{ $\bm{d_H}$\bfseries (nm) }&\multirow{2}{*}{ $\bm{d_L}$\bfseries (nm) }& \bfseries Total \\
&&&&&& \bfseries thickness ($\bm{\mu}$m) \\
\hline
(\emph{HL})\textsuperscript{10}&\multirow{7}{*}{0.7}&\multirow{7}{*}{0.5}&1787&286.96&209.18&4.96\\
(\emph{HL})\textsuperscript{20}&{}&{}&1724.5&276.81&201.7&9.57\\
(\emph{HL})\textsuperscript{30}&{}&{}&1712.5&274.86&200.26&14.25\\
(\emph{HL})\textsuperscript{40}&{}&{}&1708.5&274.21&199.78&18.96\\
(\emph{HL})\textsuperscript{50}&{}&{}&1706.5&273.88&199.55&23.67\\
(\emph{HL})\textsuperscript{75}&{}&{}&1705&273.64&199.37&35.48\\
(\emph{HL})\textsuperscript{100}&{}&{}&1704&273.48&199.25&47.27\\
\hline
\emph{H}\textsuperscript{80}&0.6&---&1721&234.04&---&18.72\\
\hline
\end{tabular}
\end{table*}

\begin{figure*}[bt]
\centering
\includegraphics[width=1\textwidth]{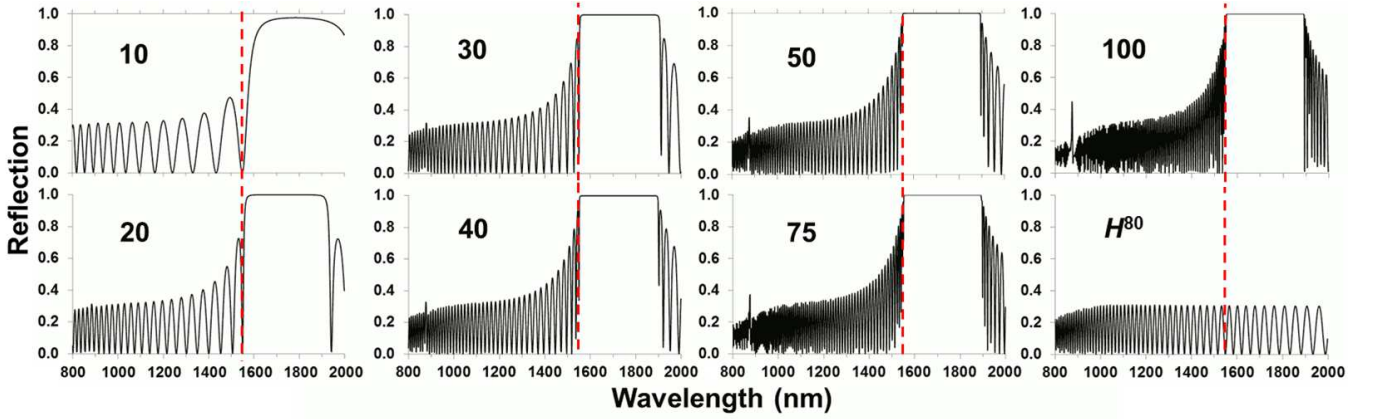}
\caption{Simulated reflectivity spectra corresponding to the DBR designs in Table~\ref{tab: MLDP}. The dashed red lines indicate the 1550-nm point.} 
\label{fig: AllRefs}
\end{figure*}

The simulated normal-incidence reflectivity characteristics of all structures included in Table~\ref{tab: MLDP} are shown in Fig.~\ref{fig: AllRefs}. The 1550-nm point is indicated by the dashed red vertical lines. It is immediately evident that increasing the number of bilayers results in a sharper stop band edge and an increased density of Fabry-P\'erot fringes on either side of the stop band. In contrast, the reflectivity spectrum of $H^{80}$ shows that the structure behaves like a thick film with no apparent photonic band gap, but with an anti-reflection characteristic at approximately 1550\,nm. 

\subsection{Effective group refractive index calculation}
\begin{figure}[htbp]
\centering
\includegraphics[width=0.45\textwidth]{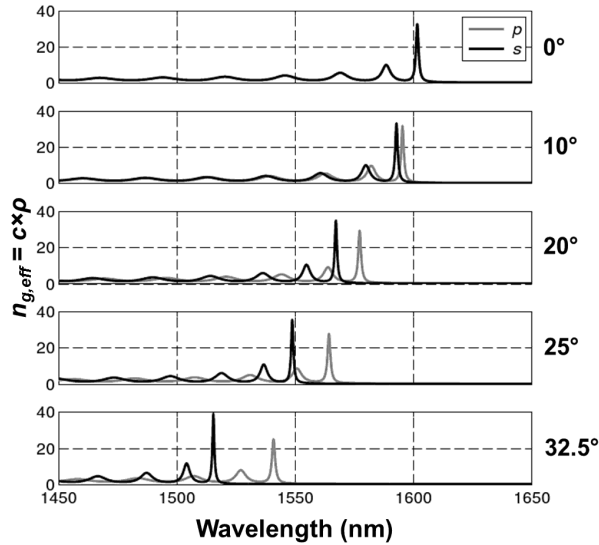}
\caption{Variation of the effective group index $n_\mathit{g,eff}$ with incident wavelength and angle for both electromagnetic polarizations in a 40-bilayer structure.}
\label{fig: neffvsAngle}
\end{figure}

The band structure of a PC may be expressed as a \emph{dispersion relation} between the frequency of the mode $\omega$ and its Bloch wave vector $K$~\cite{Yeh1988,Gutman2009}. In a periodic medium the Bloch wave vector may take any value taken by the free-space wave vector 
\begin{equation}
k=\frac{2\pi}{\lambda}=\frac{n\omega}{c}
\label{eq: Bloch5}
\end{equation}
but the two are not necessarily the same. The $\omega(K)$ of a structure determines the phase and group velocity of EM waves for a particular frequency and are respectively given by
\begin{eqnarray}
\left.v_p\right|_\omega &=& \frac{\omega}{K};\nonumber \\
{}\nonumber\\
\left.v_g\right|_\omega  &=& \frac{d\omega}{dK}.
\label{eq: Bloch6}
\end{eqnarray}
The phase velocity describes the average motion of a single point on an EM wave front, while the group velocity may be thought of as tracking the propagation of the energy carried by the wave. In homogeneous bulk material, the two are identical and are described by a single index of refraction; within periodic materials, this identity does not necessarily hold. This suggests the definition of ``effective'' indices of refraction that refer to wave propagation in periodic material:
\begin{eqnarray}
n_\mathit{p,eff}=\frac{c}{v_p};\nonumber \\
{}\nonumber \\
n_\mathit{g,eff}=\frac{c}{v_g}.
\end{eqnarray}
The photonic density of states (DOS) is defined as
\begin{equation}
\rho_\gamma(\omega)=\frac{dk}{d\omega};
\end{equation}
given that this is simply the reciprocal of Eq.~\ref{eq: Bloch6}, it is apparent that
\begin{equation}
n_\mathit{g,eff}=c\times\rho_\gamma.
\end{equation}
As shown by Bendickson\etal~\cite{Bendickson1996}, the photonic DOS may be calculated for an arbitrary, finite multilayer stack utilizing output from a transfer-matrix model. The transmission coefficient $t$ is a complex number that may be expressed as $x+iy$. Letting $x'=dx/d\omega$ and $y'=dy/d\omega$, $\rho_\gamma$ may be found:
\begin{equation}
\rho_\gamma(\omega)=\frac{1}{D}\frac{y'x-x'y}{x^2+y^2},
\end{equation}
where $D$ is the total thickness of the structure. This expression is true for either $s$ or $p$ polarization. The photonic DOS was calculated for a 40-bilayer structure with layer porosities 0.7 and 0.5 and thicknesses 283.22\,nm and 206.43\,nm, respectively, to give a photonic band gap center wavelength of 1764\,nm at normal incidence. This was done for both polarizations for a varying angle of incidence; the results are shown in terms of the effective group index in Fig.~\ref{fig: neffvsAngle}.

The figure illustrates the strong dependence of the photonic properties of the structure on the incident angle. Both regions of low group velocity/high $n_\mathit{g,eff}$ are shown to shift toward shorter wavelengths as the incident angle increases. However, the characteristic is distinct for the two polarizations due to the one-dimensional discrete translational symmetry of the structure. In effect, as the incident angle changes, the electric and magnetic fields must satisfy different boundary conditions for the two polarizations, whereas for normal incidence the two polarizations are indistinct. The features in the $s$-polarized characteristic are shown to shift more drastically than those in the $p$-polarized curve as the angle changes; in addition, the peak value of $n_\mathit{g,eff}$ for $s$-polarized light slightly \emph{increases} for a steeper angle, while the maximum value for $p$-polarized light shows a marked \emph{decrease}. 

\subsection{Indirect calculation of potential absorption enhancement}
\label{sec: ICPAE}
\begin{figure}[htbp]
\centering

\includegraphics[width=0.45\textwidth]{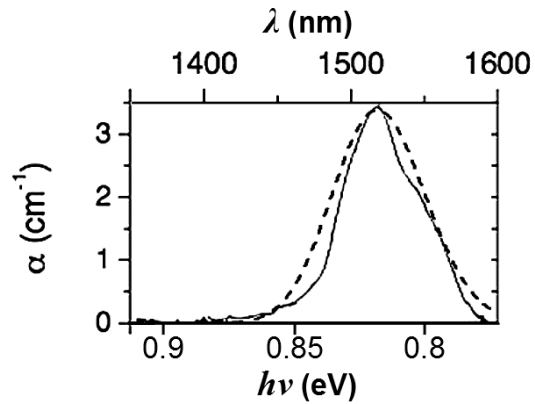}
\caption{Gaussian fit (dashed line) of measured NaYF$_4$:\erbium \enlev{4}{I}{15}{2}$\rightarrow$\enlev{4}{I}{13}{2} absorption coefficient (solid line) for simulation of absorption enhancement within multilayer structure. Measured data were taken from~\cite{Shalav2006}.}
\label{fig: GaussFit}
\end{figure}

The simulations above demonstrate that an enhanced DOS arises within periodic structures due to the suppression of the group velocity of electromagnetic wave propagation. In order to examine the impact of these modes on real \erbium absorption, calculations were carried out employing a Gaussian approximation to the NaYF$_4$:\erbium \enlev{4}{I}{15}{2}$\rightarrow$\enlev{4}{I}{13}{2} absorption coefficient as determined by Shalav~\cite{Shalav2006}; the real and approximated spectra are shown in Fig.~\ref{fig: GaussFit}. The Gaussian is given for wavelength $\lambda$ by
\begin{equation}
\alpha_\mathrm{\erbium}(\lambda) = \alpha_\mathrm{\erbium,0} \exp{\left[-\left(\frac{\lambda-\lambda_0}{W}\right)^2\right]}
\end{equation} 
where $\lambda_0$ is the central (peak) wavelength, $\alpha_\mathrm{\erbium,0}=\alpha_\mathrm{\erbium}(\lambda_0)$ and $W$ is the standard deviation from $\lambda_0$. The Gaussian curve shown here has $\alpha_\mathrm{\erbium,0}$=3.4\,cm$^{-1}$, $\lambda_0$=1518\,nm and $W$=30\,nm\footnote{Because microcrystalline NaYF$_4$:\erbium demonstrates diffuse scattering, the Kubelka-Munk theory has been employed elsewhere~\cite{Fischer2012a} to give a peak $\alpha$ of almost 6\,cm$^{-1}$ for these phosphors. Since the purpose of this analysis is to calculate enhancement for an arbitrary UC material embedded in a photonic crystal, an exact $\alpha$ is not necessary.}.

This approximation allows for the simulation of a PSi:\erbium material by the transfer matrix method as suggested by Bisson and Ueda~\cite{Bisson2011}. A long-wavelength imaginary component $\kappa_\mathrm{\erbium}\xspace$ may be added to the Bruggeman complex index of refraction of PSi:
\begin{eqnarray}
\centering
\tilde{n}_\mathrm{PSi:\erbium}(\lambda)&=&\tilde{n}_\mathrm{PSi}(\lambda,P)+iP\kappa_\mathrm{\erbium}(\lambda)\nonumber\\
{}\nonumber \\
&=&\tilde{n}_\mathrm{PSi}(\lambda,P)+iP\frac{\lambda\alpha(\lambda)}{4\pi}.
\end{eqnarray}
The prefactor $P$ on $\kappa_\mathrm{\erbium}$ implies that the total absorption in \erbium is linearly proportional to the porosity of the layer in which it is deposited. This dependence is \emph{positive} since for a greater porosity a greater \erbium concentration is possible in principle (due to the increasing specific surface area).

Based on these assumptions, the \erbium absorption in an arbitrary multilayer PSi:\erbium stack may be estimated by a transfer matrix calculation of the reflectance $R(\lambda)$ and transmittance $T(\lambda)$ for the stack as described above; the absorption is then given by 
\begin{equation} 
A(\lambda)=1-R(\lambda)-T(\lambda).
\end{equation}

In order to examine the impact of exciting \erbium ions within band-edge modes of PSi DBRs, the eight multilayer structures from Table~\ref{tab: MLDP} were again simulated by the transfer matrix approach, but modified as described here for the case of \erbium doping. 

\begin{figure}[bt]
\centering
\includegraphics[width=0.5\textwidth]{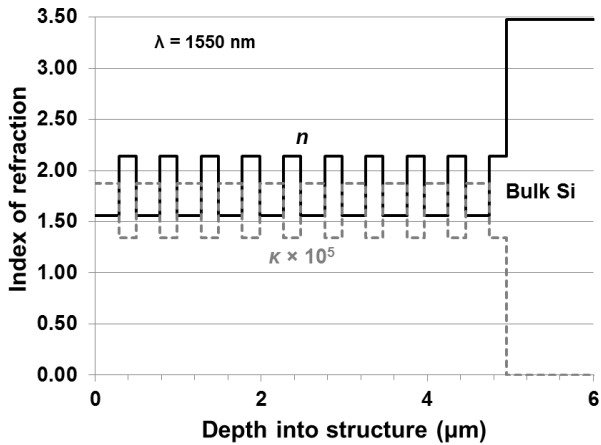}
\caption{Refractive index profile of (\emph{HL})$^{10}$ for $\lambda$=1550\,nm, showing the real (solid black line) and imaginary (dashed grey line) components. The imaginary axis has been scaled up by a factor of 10$^5$. The zero-depth point corresponds to the PSi/air interface; the PSi/bulk Si interface is shown.}
\label{fig: RefIndexStruc}
\end{figure}

Figure~\ref{fig: RefIndexStruc} shows the refractive index profile of structure (\emph{HL})$^{10}$ for $\lambda$=1550\,nm, demonstrating periodicity of the real (solid black line) and imaginary (dashed grey line) parts of $\tilde{n}_\mathrm{PSi:\erbium}$. Because Si itself is not absorbing at this wavelength, a finite $\kappa_\mathrm{PSi:\erbium}$ is due entirely to \erbium and is positively correlated with $P$, in contrast to $n_\mathrm{PSi:\erbium}$ as shown. Note that the imaginary index has been magnified by a factor of 10$^{5}$. The real part of the refractive index of \erbium is presumed to have negligible impact on the optical characteristics of the structure.

\begin{figure*}[hbt]
\centering
\includegraphics[width=1\textwidth]{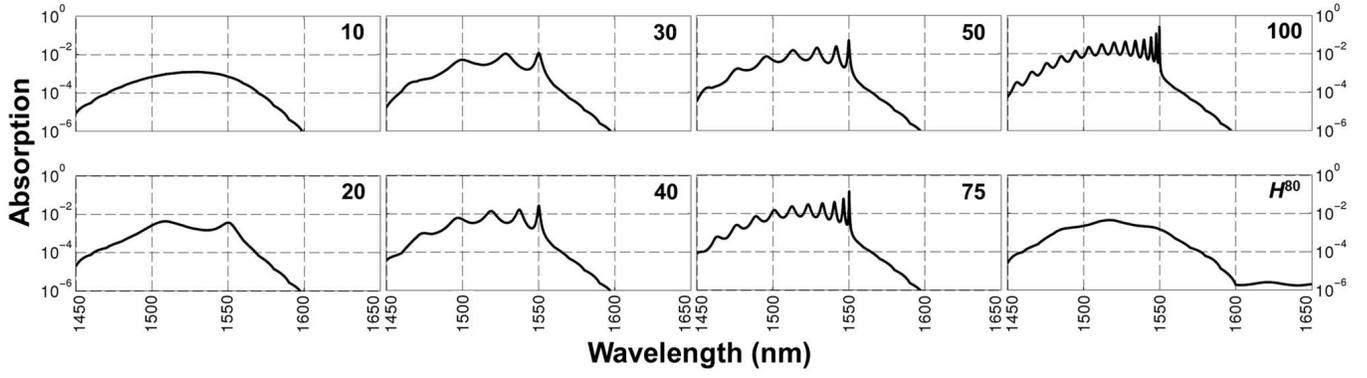}
\caption{Indirectly-calculated absorption in all structures from Table~\ref{tab: MLDP} (plotted on a logarithmic scale) for simulated \erbium profile. Bulk-like absorption is shown for the $N$=10 and \emph{H}$^{80}$ structures, with the strongly periodic structures showing modified, enhanced absorption spectra outside the photonic band gap (i.e., $\lambda<$1550\,nm).}
\label{fig: AbsSpecsAll} 
\end{figure*}

\begin{figure*}[hbt]
\includegraphics[width=\textwidth]{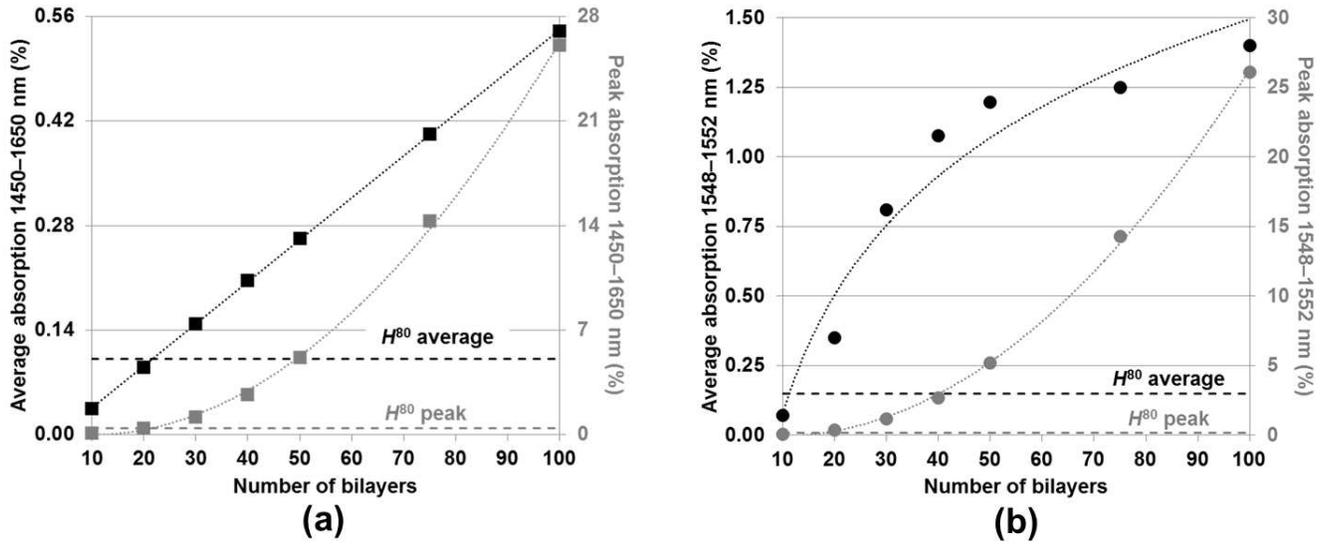}
\caption{Average (black symbols, left axis) and peak (grey symbols, right axis) absorption as a function of the number of high- and low-porosity bilayers in the DBR. Horizontal dashed lines show results for the undifferentiated \emph{H}$^{80}$ structure. Fine dotted lines are guides to the eye. \textbf{(a)}~Results for the entire band between 1450 and 1650\,nm. \textbf{(b)}~Results for narrow band around 1550\,nm.}  
\label{fig: AbsEn}
\end{figure*}

Figure~\ref{fig: AbsSpecsAll} presents the results of the absorption calculation for all structures for the spectral range between 1450 and 1650\,nm. The strengthening of the photonic properties of the structure has an obvious impact on the \erbium absorption: although the reflectivity of the structure is effectively zero at 1550\,nm in all cases, absorption is seen to dramatically increase \emph{at all wavelengths} in this range outside the photonic band gap as the number of bilayers increases, consistent with the apparent increase in $n_\mathit{g,eff}$ far from the band edge for (\emph{HL})$^{40}$ as shown in Fig.~\ref{fig: neffvsAngle}. The lobe-like features that owe to higher-order enhancement peaks are clearly ``imprinted'' upon the absorption spectra, except in the $N$=10 and $H^{80}$ cases, which approximate the ``bulk'' scenario. 
To clarify the enhancement in absorption as a function of the photonic structure, the average and peak absorption over the 1450--1650-nm spectral region are plotted against the number of bilayers in Fig.~\ref{fig: AbsEn}. The corresponding values for $H^{80}$ are plotted for comparison as horizontal dashed lines.

The figures show that both the average and the peak absorption increase with an increasing number of bilayers in both the broad- and narrow-band cases. The peak absorption is similar in both cases because the narrow band of Fig.~\ref{fig: AbsEn}(b) was chosen to coincide with the band-edge absorption peak. Both results show that both the average and peak absorption in the periodic structure are enhanced over the quasi-bulk \emph{H}$^{80}$ case when the number of bilayers exceeds 20, although the (\emph{HL})$^{20}$ structure is only half as thick as \emph{H}$^{80}$.

\begin{figure*}[bt]
\includegraphics[width=1\textwidth]{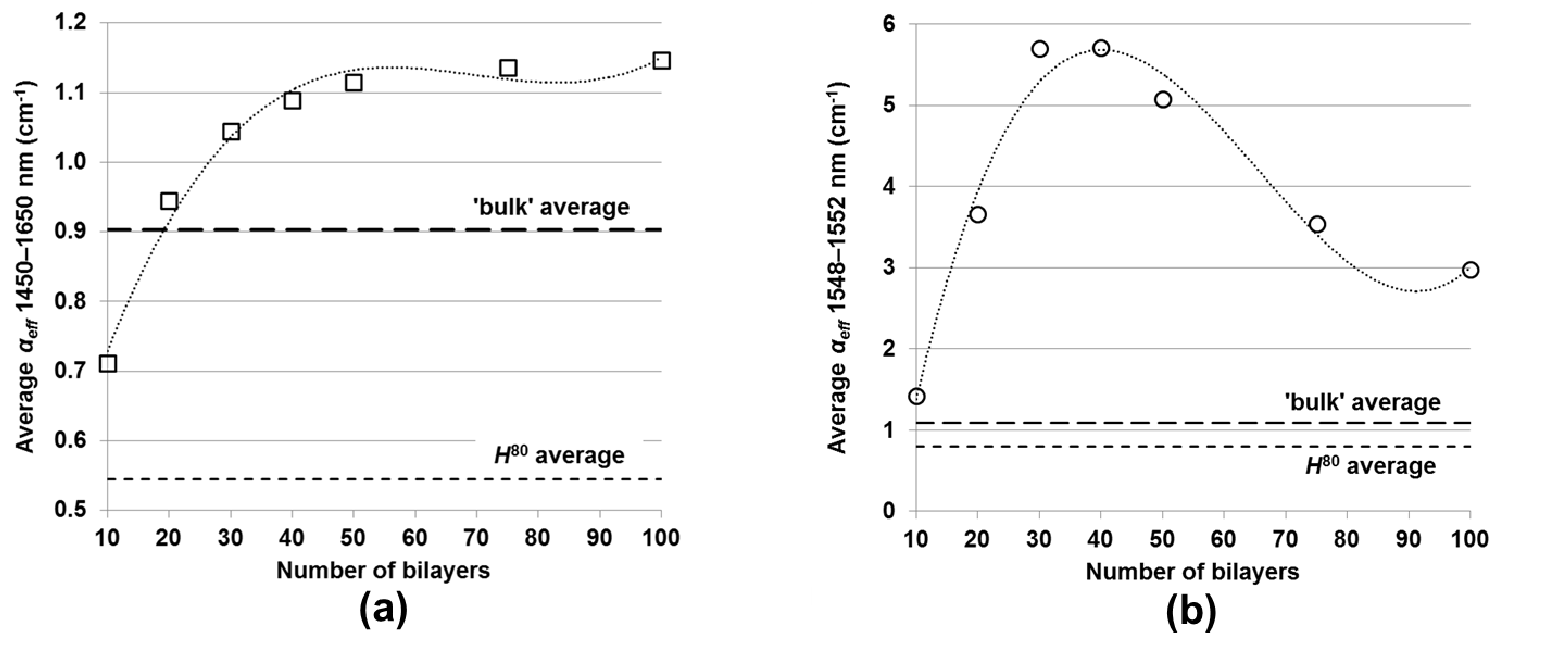}
\caption{Effective absorption coefficient (Eq.~\ref{eq: EffAbsCoeff}) as a function of the number of DBR bilayers. Wide horizontal dashes indicate the result for a bulk phosphor and narrow dashes correspond to the undifferentiated $H^{80}$ case. Fine dotted lines are guides to the eye. \textbf{(a)}~Broad-band results. \textbf{(b)}~Narrow-band results.}  
\label{fig: AbsEff}
\end{figure*}

Structure thickness is expected to play an important role in total absorption, since a thicker structure contains a greater volume of \erbium. Because the total thickness increases by an order of magnitude over the range of structures examined here, an ``effective absorption coefficient'' $\alpha_\mathit{eff}$ was calculated for each structure by the expression
\begin{equation}
\alpha_\mathit{eff}=\frac{\ln{(1-A')}}{D},
\label{eq: EffAbsCoeff}
\end{equation}
where $A'$ is the average absorption and $D$ is the structure thickness in each case. The results are shown in Fig.~\ref{fig: AbsEff}; they illustrate that the absorption enhancement over the bulk case is not merely due to the increased volume of \erbium contained in thicker structures. As demonstrated in Fig.~\ref{fig: AbsEff}(a), the average broad-band (i.e., 1450--1650-nm) response is enhanced in all cases, ranging from a 41.1\% increase over the $H^{80}$ result for (\emph{HL})$^{10}$ to a 122.9\% improvement in the case of 100 bilayers. The case for a bulk phosphor (not contained within a porous structure) is also shown; as expected, its absorption is higher than that of $H^{80}$, but photonic structures with $N\geq 20$ still fare better than bulk material. The narrow-band results in Fig.~\ref{fig: AbsEff}(b) display more significant enhancement in the immediate vicinity of the band edge, particularly for medium-thickness structures ($N$=30--50). However, these high enhancements have an increasingly marginal impact on the overall absorption because of the vanishing width of the enhancement peaks, as demonstrated in Fig.~\ref{fig: AbsSpecsAll} above.

\subsection{Ideal positioning of the photonic band gap}
The results above clearly demonstrate the fundamental utility of a slow-light structure: for the chosen material properties, it is possible for the effective absorption in an optical material to exceed that of the ordinary bulk case even when the slow-light behavior is weak (e.g., for $N$=20). Importantly, this enhancement is a broad-band effect although it is particularly strong near the photonic band edge. As such the total broad-band absorption enhancement is related to both the absorption spectral lineshape of the absorbing medium and the spectrally-resolved effective group index of the photonic crystal, i.e., the position of the photonic band gap. 

\begin{figure*}[bt]
\includegraphics[width=0.8\textwidth]{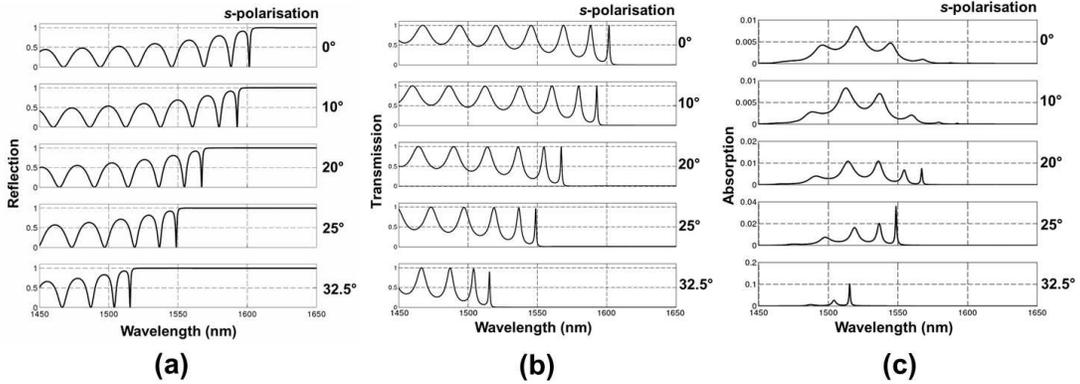}
\caption{Simulated angular dependence of the \textbf{(a)}~reflection, \textbf{(b)}~transmission and \textbf{(c)}~absorption spectra for an \erbium-doped 40-bilayer structure for $s$-polarized light.}
\label{fig: AngDepspol}
\end{figure*}

\begin{figure*}[bt]
\includegraphics[width=0.8\textwidth]{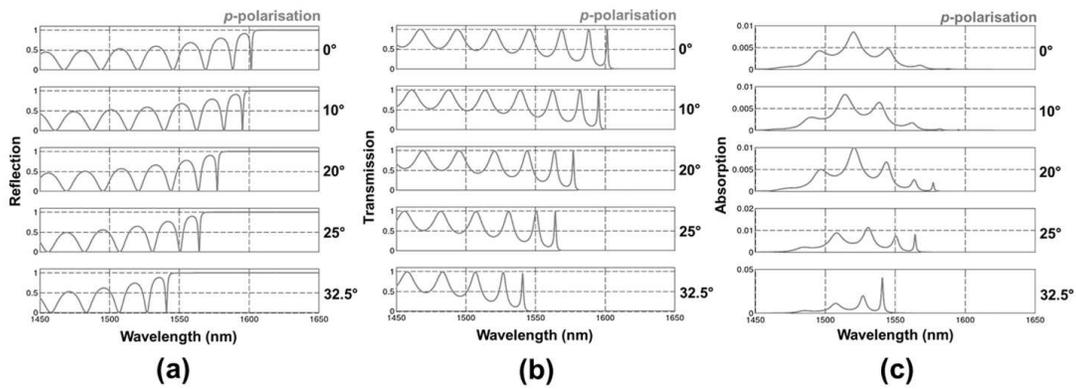}
\caption{Simulated angular dependence of the \textbf{(a)}~reflection, \textbf{(b)}~transmission and \textbf{(c)}~absorption spectra for an \erbium-doped 40-bilayer structure for $p$-polarized light.}
\label{fig: AngDepppol}
\end{figure*}

\begin{figure}[tb]
\centering
\includegraphics[width=0.5\textwidth]{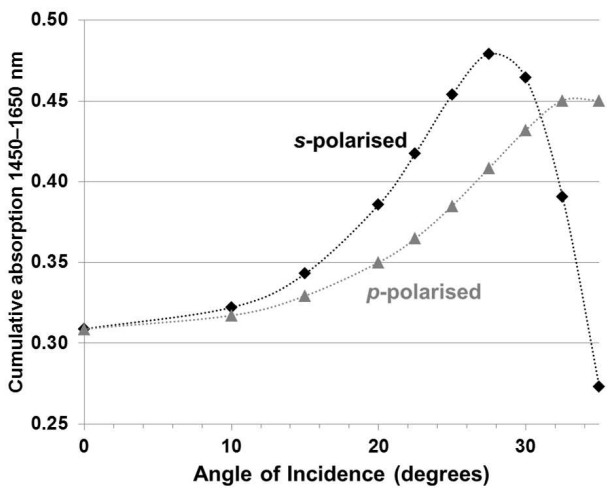}
\caption{Simulated broad-band cumulative absorption in an \erbium-doped 40-bilayer structure for both polarizations as a function of the angle of incidence.}
\label{fig: AngDepSumm}
\end{figure}

Figures~\ref{fig: AngDepspol}--\ref{fig: AngDepSumm} illustrate the simulated interaction between the photonic properties of the 40-bilayer DBR used for DOS calculations above (Fig.~\ref{fig: neffvsAngle}) and the \erbium itself. In Fig.~\ref{fig: AngDepspol}, the position of the photonic band edge is effectively varied by changing the angle of incidence for $s$-polarized light. As shown in Fig.~\ref{fig: neffvsAngle} above, as the incident angle deviates from normal incidence, the spectral features of the PC shift toward shorter wavelengths. Varying the incident angle as opposed to the structural or material properties of the DBR simplifies the assessment of optical properties both in simulation and in experiment. Figure~\ref{fig: AngDepspol} reiterates the correspondence between the reflection and transmission characteristics of the structure as the incident angle varies. In addition, it demonstrates that the absorption spectrum of the DBR---essentially a modified and enhanced \erbium absorption spectrum---changes considerably over this range. Peak absorption is seen to exceed 10\% for an angle of 32.5\dgr, at which the photonic band edge is positioned near the intrinsic \erbium absorption peak (see Fig.~\ref{fig: GaussFit}).

A similar result is seen for $p$-polarized light in Fig.~\ref{fig: AngDepppol}. However, over this range of incident angles the photonic band edge does not approach the intrinsic \erbium absorption peak ($\sim$1518\,nm) and the peak absorption is therefore maintained below 5\%.

A more useful measure of the absorption enhancement in the broad-band case is the \emph{cumulative} absorption over this spectral range, found by integrating the area under the absorption curve for both polarizations. The result is shown in Fig.~\ref{fig: AngDepSumm} for a finer spacing of angles than is shown in Figs.~\ref{fig: AngDepspol} and \ref{fig: AngDepppol}. For $s$-polarized light, the results clearly show that maximum absorption over this spectral region does not necessarily correspond to the coincidence of the band-edge field enhancement peak and the intrinsic \erbium absorption peak. When the band edge is positioned within the long-wavelength ``tail'' of the intrinsic \erbium absorption spectrum, the entire absorption spectrum may contribute to the cumulative absorption over this spectral region. When the photonic band edge is pushed to shorter wavelengths to coincide with the intrinsic \erbium absorption peak, absorption in the long-wavelength \erbium tail is suppressed by the photonic band gap and the cumulative absorption plummets. 

For $p$-polarized light, the situation is not as clear since the photonic band edge does not demonstrate a blue shift as drastic as that seen for $s$ polarization. However, the characteristic appears to roll off between 32.5\dgr and 35\dgr, implying again that a maximum cumulative absorption is reached when the band edge peak is at a longer wavelength than the intrinsic \erbium absorption peak at 1518\,nm.

It bears noting that it is possible to excite optical oscillators within slow light modes on either side of the fundamental DBR photonic band gap. The simulations presented above have only examined the short-wavelength/high-frequency band edge modes. In relative terms, these modes confine themselves to regions of low refractive index, which are presumed to contain to higher concentrations of \erbium in our structures. Moreover, the presence of the photonic stop band at wavelengths slightly longer than the region of enhanced absorption may present the additional benefit of suppressing radiative \emph{emission} from the first excited state. The peak emission wavelength is expected to be red-shifted from the peak absorption wavelength by an amount given by the McCumber relation~\cite{Bisson2011}; the presence of a photonic stop band may inhibit this emission while simultaneously enhancing absorption. In contrast, a stop band at a shorter wavelength than the absorption peak is likely to compete with \erbium absorption.

\section{Protoyping of Erbium-Doped Porous Silicon Slow-Light Structures}
\label{ch: SlowLightExp}

The previous section demonstrated that a substantial increase in the \erbium absorption coefficient---in excess of the bulk value---may be produced in periodic slow-light structures. Detailed balance analysis invoking realistic \erbium and solar cell parameters has been used by the present authors~\cite{Johnson2012a} to illustrate the relationship between optical absorption within the UC layer and UC-PV device efficiency. That analysis emphasizes that very strong absorption---i.e., either a high absorption coefficient or a very thick UC layer---is needed to produce an efficiency increase over the single-junction limit in a c-Si-based UC-PV device.

\begin{figure}[bt]
\includegraphics[width=0.5\textwidth]{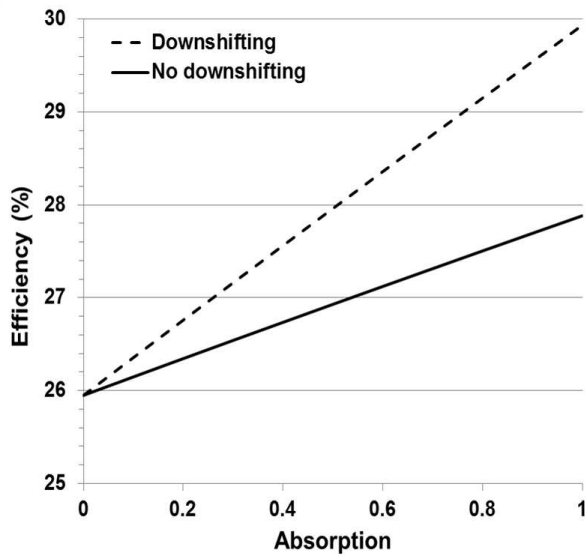}
\caption{Detailed balance limiting efficiency for a realistic c-Si solar cell coupled to an \erbium-like UC layer~\cite{Johnson2012a} as a function of absorption in the UC layer with (dashed line) and without (solid line) a downshifting sensitizer.}
\label{fig: RealErEff}
\end{figure}

Figure~\ref{fig: RealErEff} demonstrates the one-sun limiting efficiency of a realistic c-Si solar cell coupled to a generalized \erbium-like UC layer~\cite{Johnson2012a}. The limiting efficiency of the device is shown as a function of the average absorption in the UC layer. The zero-absorption point corresponds to the single-junction limit for the generalized non-ideal c-Si solar cell (25.95\%). The solid line indicates the efficiency of a UC-PV device based on the ``intrinsic'' absorption bandwidth of the \enlev{4}{I}{15}{2}$\rightarrow$\enlev{4}{I}{13}{2} \erbium transition, taken to be 1450--1650\,nm. This trend shows that only a slight increase in device performance ($\sim2\%$) is achievable even for total absorption over this spectral band. When a sensitizing element is added to ``downshift'' photon energy from 1180--1450\,nm into the \erbium absorption region~\cite{Loper2007,Goldschmidt2008}, the absolute efficiency gain approaches 4\% for full absorption.

Unfortunately, based on this model the average absorption across this spectral region is very small for the \erbium-doped DBR, less than 1\% regardless of the presence of a photonic structure. The peak absorption may be much higher, approaching 30\% near 1550\,nm for normal incidence for 100 bilayers.

For these reasons, the proposed slow-light structure alone is not expected to serve as a sufficient basis for viable c-Si UC-PV, despite its apparently strong broad-band performance. However, it may demonstrate useful narrow-band absorption enhancement behavior that allows it to serve as a central component of a high-efficiency UC layer. Such a structure also serves as a useful prototype for studying the slow-light enhancement of UC luminescence. Luminescent PSi:\erbium was therefore used to fabricate a prototype slow light structure. 

\subsection{Temperature and doping-induced degradation of photonic structures}
\begin{figure}[bt]
\centering
\includegraphics[width=0.5\textwidth]{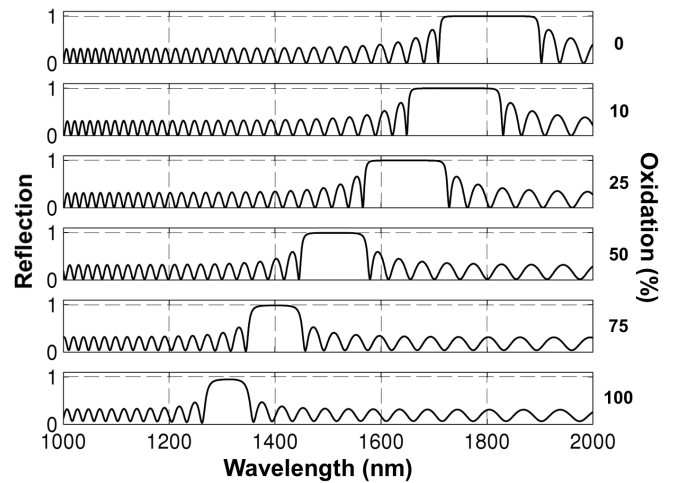}
\caption{Calculated shift in the photonic stop band of a 40-bilayer PSi DBR (with layer porosities 0.6 and 0.5 and thicknesses 245\,nm and 210.74\,nm, respectively) as the extent of oxidation increases from 0 to 100\%.}
\label{fig: DBRRefOx}
\end{figure}

The design of an \erbium-doped PSi DBR structure with specified optical characteristics is complicated by the required high-temperature oxidation and densification annealing steps required to activate the \erbium ions present~\cite{Lopez2000,Elhouichet2007}. While the alteration of the chemical environment of the \erbium ions within the structure is ultimately desirable, oxidation of PSi results in a lowering of its refractive index; heating simultaneously induces structural changes. Both of these effects must be anticipated in the initial design of the multilayer structure, implying that the processes must be to some extent predictable. The alteration of PSi DBR optical characteristics as a function of oxidation has been previously examined~\cite{Lopez2000a,Charrier2012}; the reduction in the refractive index with oxidation time and temperature produces a blue shift in the DBR center wavelength. Figure~\ref{fig: DBRRefOx} demonstrates the shift in the calculated optical properties of a PSi DBR as the extent of oxidation increases from 0 to 100\%. In accordance with Eq.~\ref{eq: DBRwidth}, the the photonic stop band narrows as the refractive index contrast in the structure is decreased. 

\begin{figure}
\centering
\includegraphics[width=0.5\textwidth]{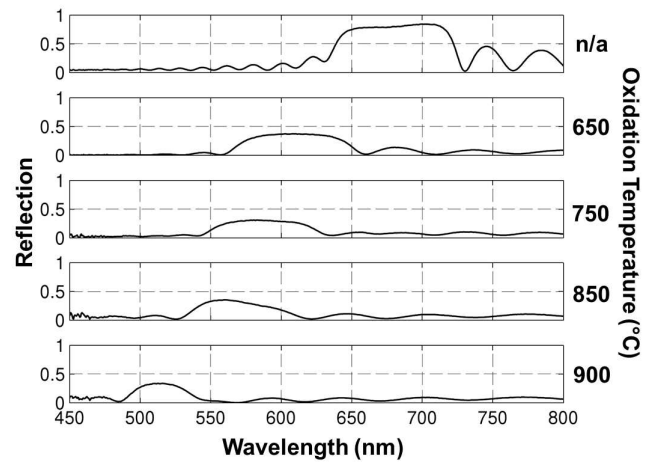}
\caption[Evolution of the visible-wavelength reflectivity spectrum of a 20-bilayer \erbium-doped PSi DBR with annealing temperature]{Evolution of the visible-wavelength reflectivity spectrum of a 20-bilayer \erbium-doped PSi DBR with annealing temperature. Samples were annealed for 30\,min in air at the temperatures indicated; the sample labelled ``n/a'' is as-doped without annealing. A temperature-dependent blue shift of the high-reflectivity band is apparent due to the reduction of the refractive index of the PSi skeleton by oxidation. In addition, substantial degradation of the photonic properties is evident, indicating that annealing also imparts significant structural incoherence.}
\label{fig: DBR55RefOx}
\end{figure}

The simple progression shown in Fig.~\ref{fig: DBRRefOx} is not seen for very thick \erbium-doped structures annealed at 1100\dgr C, however. In those cases the structural changes induced by the doping and high-temperature steps result in damage to the skeleton, layer densification, sintering~\cite{Muller2003} and interface roughening~\cite{Charrier2012} in addition to the expected variation due to oxidation. Figure~\ref{fig: DBR55RefOx} shows a case of drastic degradation of a photonic structure as the extent of oxidation increases. 
The 20-bilayer DBR was prepared by etching a 500-$\mu$m-thick \hkl<100> $p$-type (B-doped) c-Si wafer with a nominal resistivity of 60--70\,m$\Omega$-cm. Etching was carried out at room temperature in 25\% HF solution.

Etching currents of 21\,mA and 90\,mA were selected to give layer porosities of 0.5 and 0.7, respectively. Etching times were 4.630\,s and 1.497\,s, respectively, for layer thicknesses of 76.724\,nm and 108.67\,nm. This DBR was designed to produce a reflectivity band at visible wavelengths, as is shown in Fig.~\ref{fig: DBR55RefOx}. The DBR was electrochemically doped for 60\,min at 200\,$\mu$A, then cleaned by the RCA-3/SC-2 process~\cite{Lopez1998}. Figure~\ref{fig: DBR55RefOx} shows the visible-wavelength normal-incidence reflectivity profiles for sections of the DBR annealed in air for 30\,min at various temperatures (650, 750, 850 and 900\dgr C) along with that of an unannealed portion. The unannealed portion demonstrates an incomplete reflectivity characteristic, indicating some disorder and imperfections are present prior to annealing. It is apparent that along with a blue shift of the reflectivity peak, the photonic structure is almost completely degraded by oxidation and structural modification as a result of the high-temperature step. In particular, the degradation is characterized by a lowering of the peak reflectivity; the reflectivity band edges resultantly become less steep. It was found that the three-part Bruggeman model of structure oxidation cannot be used to account for the demonstrated alteration of the photonic structure.

As shown in~\cite{Johnson2011}, the magnitude of the slow-light enhancement is to some extent proportional to the ``steepness'' of the photonic band edge; the degradation of the stop band due to high-temperature structural modification clearly presents a barrier to the use of this type of structure as a photonic crystal. All real photonic crystals are subject to disorder to a certain extent due to  variations in fabrication. Disorder in periodic structures intuitively introduces the potential for broadening of interference features, undermining the intended operation of the structure, e.g., optical confinement or other resonant behavior~\cite{Greshnov2007,Greshnov2008,Rybin2009}. 

\subsection{Fabrication and optical characterization}
Attempts to fabricate a high-quality luminescent PSi:\erbium DBR with a band-edge mode near 1550\,nm (using techniques similar to those described elsewhere~\cite{Lopez2001}) illustrated a fundamental tradeoff of the PSi:\erbium system. Although the structures appeared highly uniform and strongly reflecting prior to annealing, the high-temperature steps required for luminescence destroyed the samples or degraded their optical properties in an unpredictable fashion. Restricting annealing to low temperatures to avoid significant structural modification resulted in immeasurable \erbium luminescence. 

\begin{figure}[bt]
\centering
\includegraphics[width=0.45\textwidth]{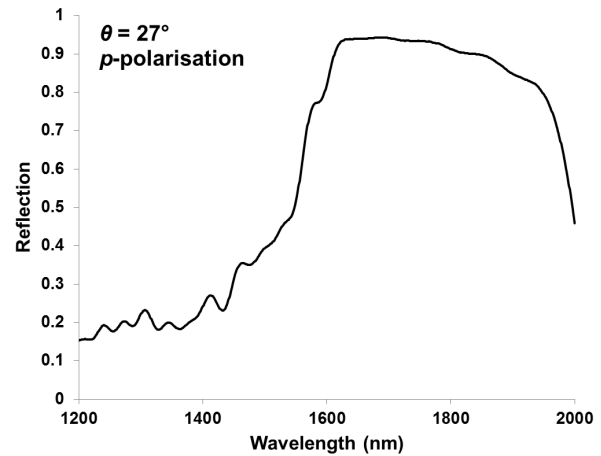}
\caption{Near-infrared $p$-polarized specular reflectivity spectrum of DBR42N, a 30-bilayer PSi:\erbium DBR, for an incident angle of 27\dgra.}
\label{fig: DBR42R27}
\end{figure}

Figure~\ref{fig: DBR42R27} shows the near-infrared reflectivity of the best prototype structure fabricated for this work. The 30-bilayer PSi:\erbium DBR (``DBR42N'') was measured for $p$-polarization at an incident angle of 27\dgra relative to normal incidence. An intermediate number of bilayers was chosen to take advantage of a substantial predicted absorption enhancement~\cite{Johnson2011} while also promoting structural stability throughout the fabrication process. Sample reflectivity was measured with the Perkin-Elmer Lambda 1050 UV/Vis/NIR spectrophotometer; the Universal Reflectance Accessory add-on was used to allow variation of the angle of incidence and detection, providing specular reflectance information for angles between 21\dgra and 38\dgra as shown in Fig.~\ref{fig: RAngles}. The instrument features a tungsten-halogen lamp and a Peltier-cooled InGaAs detector for the NIR spectrum. The spectral resolution in the NIR is 0.2\,nm and is calibrated to a nominal wavelength accuracy of $\pm$0.300\,nm. The beam is internally depolarized but a polarizing beamsplitter cube was inserted in the beam path to enforce $p$-polarization at the sample. 

DBR42N was etched from a 1.5--2-m$\Omega$-cm $p$-type (B-doped) \hkl<100> Si wafer at room temperature in a 25\% HF solution. Etching currents of 5.6\,mA and 99.75\,mA were selected to produce layers of 50\% and 70\% porosity, respectively. Layer etching times were 58.878\,s and 9.4\,s with two 3-s breaks per layer to promote layer uniformity and structural integrity. These etching times were chosen empirically by trial-and-error with the intent of fabricating a DBR with a short-wavelength band edge near 1550\,nm after doping and annealing. 

\begin{figure}[bt]
\includegraphics[width=.5\textwidth]{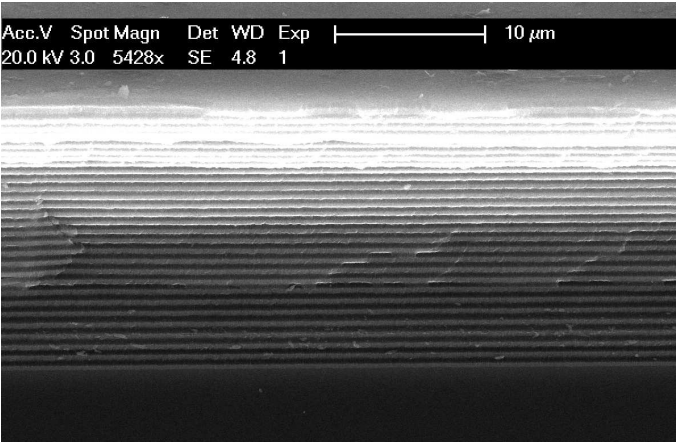}
\caption{Secondary-electron cross-sectional SEM images of DBR42N before annealing, showing regular variation of the periodic index.}
\label{fig: DBR42XSec}
\end{figure}

\begin{figure}[bt]
\centering
\includegraphics[width=0.45\textwidth]{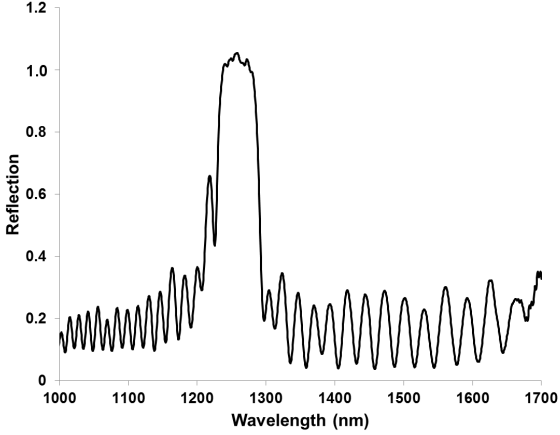}
\caption[Second-order reflectivity peak of DBR42N when measured at normal incidence prior to annealing]{Second-order reflectivity peak of DBR42N when measured at normal incidence prior to annealing, indicating layer thickness mismatch and a first-order peak centered at approximately 2500\,nm.}
\label{fig: DBR42PreVis}
\end{figure}

The cross-sectional SEM image of the pre-annealed structure in Fig.~\ref{fig: DBR42XSec} shows a high-porosity layer thickness of approximately 415\,nm and a low-porosity layer thickness of approximately 300\,nm. Simulation of such a structure gives a photonic band gap centered near 2600\,nm. While it was not feasible to measure the reflectance of this structure at long wavelengths before annealing, the NIR spectrum (shown in Fig.~\ref{fig: DBR42PreVis}) contains a second-order reflectivity band centered at 1250\,nm. This indicates the presence of a fundamental band gap centered at 2500\,nm, although the presence of a second-order peak also implies imperfect periodicity of the structure~\cite{Reece2004a}. Substantial inhomogeneity is also apparent in layers near the surface.

After etching, DBR42N was electrochemically doped with \erbium for 60\,min at 200\,$\mu$A and annealed in flowing N$_2$ (16\,mL/min) for 30\,min at 1100\dgr C. The subsequent observation of luminescence from this sample indicates that a considerable amount of oxygen remains in the system despite the lack of flowing oxygen during annealing. 

In contrast to the DBR spectra in, e.g., Fig.~\ref{fig: DBR55RefOx}, the measured spectrum of DBR42N in Fig.~\ref{fig: DBR42R27} demonstrates suppressed interference fringes in the high-transmission region at wavelengths below the photonic band gap. In addition to the reduction in periodicity mentioned above, this is a result of performing the reflectivity measurement over a large sample area (4$\times$4\,mm$^2$). While this ensures a high signal-to-noise ratio, the spot contains a considerable amount of lateral sample inhomogeneity, ``blurring'' out the interference fringes and producing a bulk-like reflectivity profile with some remnant interference features. 

Despite the degraded structural quality of the DBR, its periodicity is still sufficient to produce a strongly-reflecting region beyond 1550\,nm.
%
%\subsection{Compositional analysis}
%A Rutherford backscattering spectrometry (RBS) measurement was carried out on DBR42N using a 2-MeV \ion{He}{}{+} NEC 5SDH tandem accelerator; the results are shown in Fig.~\ref{fig: DBR42RBS}. The dots in Fig.~\ref{fig: DBR42RBS}(a) indicate measurement data, and the solid line is the simulated profile corresponding to the proposed Er distribution shown in Fig.~\ref{fig: DBR42RBS}(b). Only the first 2.5\,$\mu$m of depth are resolved, but the periodicity of the Er:Si ratio owing to the variation in layer porosity is clearly demonstrated. However, severe structural modification over this region and lateral variation in layer topology make it difficult to directly assign the features of the simulated Er profile to real structural features.
%
%\begin{figure*}[htbp]
%\includegraphics[width=0.7\textwidth]{DBR42Nc}
%\caption{Results of Rutherford backscattering measurement of DBR42N. \textbf{(a)}~Measured data (dots) and simulated spectrum (solid line). \textbf{(b)}~Proposed Er:Si profile resulting in the simulated backscattering spectrum fit to the data points in (a). The Er distribution demonstrates an obvious periodicity owing to the variation in layer density.}
%\label{fig: DBR42RBS}
%\end{figure*}

\subsection{Angular-dependent photoluminescence measurement}
\begin{figure}[bt]
\centering
\includegraphics[width=0.45\textwidth]{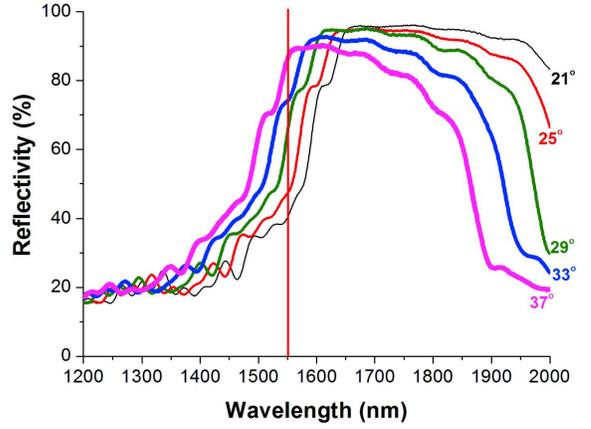}
\caption{(\emph{Color online.}) Overlayed specular $p$-polarized reflectivity spectra of DBR42N with angle of incidence and detection varying from 21\dgra to 37\dgra (in steps of 4\dgra for clarity); a monotonic blueshift in the short-wavelength band edge is evident with increasing angle. The vertical line denotes the excitation laser wavelength (1550\,nm).}
\label{fig: RAngles}
\end{figure}

As illustrated in Fig.~\ref{fig: AngDepppol} above, as the angle of incident light deviates from normal incidence (0\dgra), the features of the DBR reflectivity spectrum are expected to shift to shorter wavelengths. Figure~\ref{fig: RAngles} demonstrates this blue shift for DBR42N for incident angles ranging from 21\dgra to 37\dgra. 

Over this range of incident angles, the short-wavelength edge of the photonic stop band is shown to traverse the 1550-nm point, indicated by the vertical red line in Fig.~\ref{fig: RAngles}. It is not substantially altered otherwise (in contrast to the long-wavelength edge). This property suggests that any absorption enhancement effects related to the position of the stop band edge may be examined by minor variations in the angle of excitation, effectively simulating a range of DBR structures with a slightly different band edge position. This approach avoids the requirement of strict repeatability and controllability of structural parameters, which---as demonstrated above---is frustrated by non-uniformity induced by doping and annealing.

\begin{figure}[bt]
\centering
\includegraphics[width=0.45\textwidth]{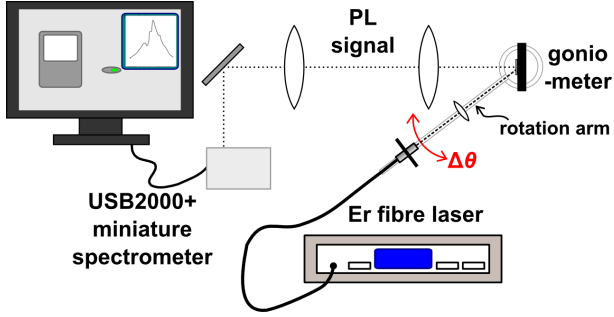}
\caption{(\emph{Color online.}) Experimental setup to achieve angular-dependent up-conversion photoluminescence measurement of DBR42N. The rotation arm sweeps out a range of angles with respect to the sample surface; UC-PL is collected normal to the sample.}
\label{fig: PLsetupAng}
\end{figure}

As shown in Fig.~\ref{fig: PLsetupAng}, room-temperature angle-dependent PL spectra were collected normal to the DBR through a 4-\textit{f} correlator telescope optical arrangement with the OceanOptics USB2000+ miniature spectrometer. The excitation source was the IPG Photonics Er fiber laser (200\,mW, $\lambda_\textit{ex}$=1550\,nm) mounted on a rotation arm and focused with a one-inch lens with a focal length of 12.5\,cm. The rotation of the arm was fixed to a goniometer beneath the sample and its axis was held coplanar with the sample surface. Thus as the arm swept through the range of incident angles, the laser spot was fixed at a single point on the sample surface. The fiber was rotated to produce \textit{p}-polarized excitation. 

\begin{figure}[bt]
\includegraphics[width=0.45\textwidth]{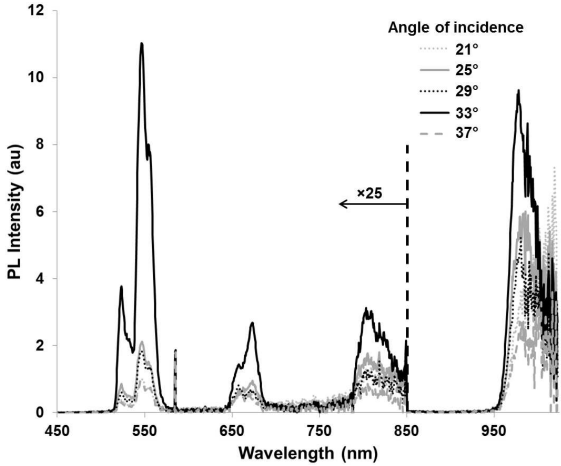}
\caption{DBR42N UC-PL spectra for various angles of incidence corresponding to the reflectivity profiles in Fig.~\ref{fig: RAngles}. Note the short-wavelength peaks have been scaled by 25$\times$.}
\label{fig: PLoverlay5}
\end{figure}

Room-temperature photoluminescence was measured for an incident angle varying in steps of 1\dgra for five spots on the sample separated by approximately 0.5\,mm. PL spectra corresponding to the incident angles shown in Fig.~\ref{fig: RAngles} for a single spot are overlayed for comparison in Fig.~\ref{fig: PLoverlay5}. Each spectrum represents the average of five 2.5-s acquisitions and has been corrected for background noise. All spectra demonstrate clear features owing to UC processes in \erbium; no other strong spectral features are apparent. Notable variation in PL intensity for all peaks is observed over this range of incident angles for this spot.

\begin{figure}[bt]
\includegraphics[width=0.45\textwidth]{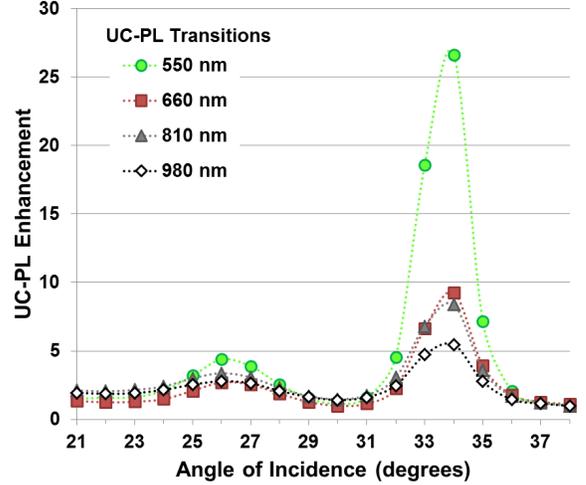}
\caption{(\emph{Color online.}) Average PL intensity enhancement for major \erbium UC transitions as a function of incident angle in DBR42N. The dotted lines are guides to the eye.}
\label{fig: gAverage}
\end{figure}

In order to assess this variation in luminescence as a function of the incident angle for all measured spots, the areas under the major peaks of the \erbium spectrum (labelled ``550\,nm'', ``660\,nm'', ``810\,nm'' and ``980\,nm'') were calculated and averaged across all spots. These intensities are then plotted together in Fig.~\ref{fig: gAverage} as a function of incident angle and normalized for each peak intensity at the angle at which it is weakest. This provides a clear demonstration of luminescence enhancement for each \erbium transition with varying angle, i.e., with varying position of the photonic band edge. A global maximum is found for all transitions at 34\dgra, with a secondary enhancement maximum apparent at 26\dgra.

\subsection{Discussion}
As the incident angle changes and the position of the photonic stop band edge shifts with respect to the laser line, the PL intensity from each major \erbium transition shows clear non-monatonic behavior that may be attributable to enhanced absorption due to photonic crystal band edge effects. In the absence of such effects, the PL trend would be determined by (1) the slight decrease in incident power density associated with an increased angle and (2) the ability of the incident field to penetrate the DBR as described by the reflectivity spectrum at each angle. Assuming a reasonable depth of focus, the spot size increases by 1.4$\times$ between angles of 21\dgra and 38\dgra with the incident power density decreasing proportionally. The impact of (1) is clearly negligible when compared to the large variation (up to 26.6$\times$) in PL intensity.

\begin{figure*}[bt]
\includegraphics[width=0.8\textwidth]{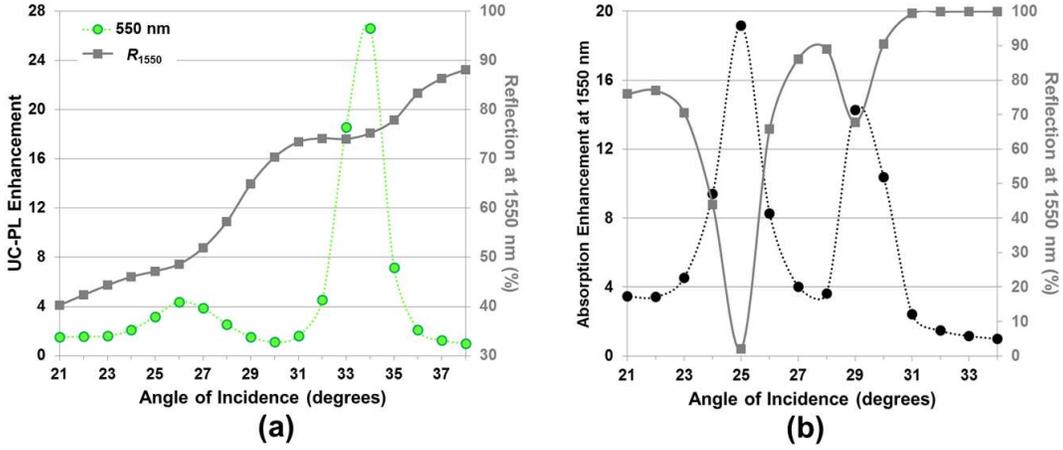}
\caption{(\emph{Color online.}) Relationship between reflectivity and absorption/UC enhancement for real and simulated DBRs: \textbf{(a)}~Enhancement trend for the 550-nm transition (green) in DBR42N against the reflectivity (grey) as measured at the laser wavelength (1550\,nm) for each angle. \textbf{(b)}~Calculated \erbium absorption (black) and reflection (grey) for 1550\,nm in the structure given in Fig.~\ref{fig: AngDepppol}. The lines in both plots are guides to the eye.}
\label{fig: RefvsPL}
\end{figure*}

Assessing the impact of (2) is straightforward. Figure~\ref{fig: RefvsPL}(a) shows the reflection from the DBR at 1550\,nm (i.e., the crossing points of the red line in Fig.~\ref{fig: RAngles} with the various reflectivity spectra). In a typical bulk case, PL intensity would peak for maximum transmission into the material. In DBR42N this condition is obtained at small angles, near ordinary dispersion conditions, and decreases as the photonic band gap is approached. However, despite the increase in reflection, a clear enhancement peak is evident for all transitions for angles near 34\dgra, where the incoupling is limited to 27\%. The implication is that the effect of pumping \erbium at certain points near a photonic band edge strongly mitigates the low incoupling efficiency of the DBR. Moreover, it is noteworthy that the two PL enhancement peaks seen near 26\dgra and 34\dgra correspond to small inflection points in the reflection/transmission behavior. 

This behavior is consistent with the simulated absorption trend as given in Fig.~\ref{fig: AngDepppol} above. Figure~\ref{fig: RefvsPL}(b)---showing data selected from Fig.~\ref{fig: AngDepppol}---presents the reflection (black line) and \erbium absorption enhancement (grey line) calculated at 1550\,nm for a 40-bilayer structure as a function of the incident angle. In this simulation the photonic band edge traverses the 1550\,nm point as the incident angle ranges from 0\dgra to 32.5\dgra. Here the absorption enhancement is taken as the calculated absorption normalized to the weakest absorption over this range (e.g., at 34\dgra). The experimental result in Fig.~\ref{fig: RefvsPL}(a) appears to demonstrate a similar correspondence between PL enhancement and reflectivity. While the reflectivity ``dips'' shown in Fig.~\ref{fig: RefvsPL}(a) are considerably shallower than those of the simulated case, it bears reiterating that the reflectivity profile shown for DBR42N is a convolution of reflectivity characteristics over a large area. Variations within the plane of the sample may have the effect of smoothing out any sharp reflectivity features. Such variation is not sampled under laser excitation, which occurs at a tightly-focused spot. The ability to simultaneously measure broad-band reflectance and photoluminescence would clarify the relationship between narrow resonances and PL enhancement. 

An additional obvious feature of the results of Fig.~\ref{fig: gAverage} is the variation of the measured peak enhancement factor between the transitions. While the angular dependence of the demonstrated enhancement is consistent for all transitions, the 550-nm transition shows an enhancement of more than 26$\times$ and the remaining lower-energy transitions show enhancements between 5$\times$ and 10$\times$. It is important to reiterate that UC luminescence from the 550- and 660-nm levels may be excited by a minimum of three 1550-nm photons. In principle, the enhancement of \erbium absorption via the \enlev{4}{I}{15}{2}$\rightarrow$\enlev{4}{I}{13}{2} transition---as postulated for an \erbium-doped slow light structure---should affect the efficiency of all transitions of the same photon number equally. A disparity might therefore be expected between the result for two- and three-photon transitions. However, the enhancement of 550-nm luminescence at 34\dgra is much larger than the 660-nm enhancement.

\begin{figure}
\includegraphics[width=0.4\textwidth]{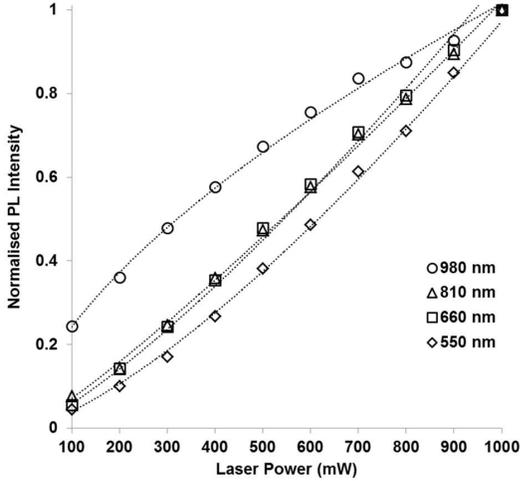}
\caption{Power-dependent UC-PL results for a single thick PSi:\erbium film with similar preparation conditions to DBR42N. The excitation wavelength was 1550\,nm; the dotted lines are exponential fits to the measured data ($R^2\geq0.997$). The 980-nm (\enlev{4}{I}{11}{2}$\rightarrow$\enlev{4}{I}{9}{2}) peak demonstrates a sub-linear dependence ($\propto P^{0.62}$) indicating saturation. The higher-energy peaks show dependencies slightly above linearity: the 810-nm peak demonstrates a $P^{1.154}$ dependence, while the 660-nm peak increases with $P^{1.257}$ and the 550-nm peak with $P^{1.373}$.}
\label{fig: CMJ021aPdep}
\end{figure}

The power-dependent UC-PL results shown in Fig.~\ref{fig: CMJ021aPdep} indicate that the luminescence of an \erbium-doped PSi sample similar to the DBR used in this study is saturated for the \enlev{4}{I}{11}{2}$\rightarrow$\enlev{4}{I}{15}{2} (980-nm) transition and approaching saturation for all other transitions. The 550-nm transition appears to be farthest from saturation conditions at the high incident powers used here, followed by the 660- and 810-nm transitions. The disparities between the magnitudes of the UC luminescence enhancements for the various transitions may reflect the distinct saturation conditions to which the transitions are subject. Further investigation is needed to clarify the source of this phenomenon.

\section{Summary and conclusions}
\subsection{Theoretical results}
Transfer-matrix-based simulations were carried out to illustrate that a high photonic DOS---due to low group velocities at particular wavelengths---may be observed within PSi DBRs, even those with an experimentally-feasible number of layers. It was suggested that such enhancement may provide a route to enhancing absorption in UC materials if the field enhancement is made to coincide with the excitation transition wavelength of \erbium. This phenomenon was modelled using an approximation to an experimental \erbium absorption lineshape (in NaYF$_4$:\erbium). It was shown that in the vicinity of the \erbium absorption peak the average effective absorption coefficient of an \erbium-doped DBR may exceed that of a bulk phosphor: $\alpha_\mathit{eff}$ may increase by more than 22\% over a broad spectral region (1450--1650\,nm, Fig.~\ref{fig: AbsEff}(a)) and by more than 400\% over a narrow region (1548--1552\,nm, Fig.~\ref{fig: AbsEff}(b)). 

The enhancement of UC luminescence itself---and therefore of solar cell efficiency enhancement---is dependent on UC dynamics and internal quantum efficiency and has not been taken into account here. As shown in Fig.~\ref{fig: RealErEff}, detailed balance calculations demonstrate that enhanced \erbium absorption may lead to increased solar cell efficiency; the required enhancement is drastic, however, even for a UC layer that demonstrates perfect internal quantum efficiency. For realistic, inefficient UC, absorption enhancement helps to mitigate poor internal quantum efficiency. A full examination of the impact of localized field enhancement on the quantum efficiency of UC dipole oscillators may be found in, e.g.,~\cite{Fischer2012}.

\subsection{Experimental results}
The experimental component of this work proposed and assessed a straightforward method for examining slow-light enhancement of optical absorption and up-conversion in real structures. Erbium-doped porous silicon DBRs were taken as prototype one-dimensional photonic crystals for this purpose, and were fabricated and characterized using an angular-dependent photoluminescence measurement. Novel microfabrication techniques have previously allowed the realization of useful structures for related investigations~\cite{ZhangF2010,Yang2011,Yang2011a}; however, while these studies allude to slow light band edge emission enhancement, it was not directly examined or observed. 

The overwhelming challenge with the approach used here was found to be the predictability and repeatability of fabrication; electrochemical doping and requisite high-temperature annealing drastically and anisotropically modify the material in a manner that degrades the optical quality of the structure. Although it was possible to produce high-quality PSi DBRs in the wavelength range of interest, it was not generally possible to dope them with \erbium and subject them to the high-temperature step without degrading or destroying the DBR.

A structure was fabricated that demonstrated optical characteristics of moderate quality with a photonic band edge near 1550\,nm (DBR42N). The dependence of the optical properties of DBRs on the angle of incidence allows the examination of particular wavelength-dependent phenomena merely by varying the angle of excitation. Photoluminescence measurements were carried out at a variety of incident angles, effectively studying the relationship between the position of the photonic band edge---where slow-light modes are expected to prevail---and absorption in \erbium ions present within the DBR. 

For this structure, non-monotonic enhancement was demonstrated as the incident angle varied between 21\dgra and 38\dgra, an angular range over which the photonic band edge traversebd 1550\,nm (the wavelength of our \erbium fiber laser). The variation in UC photoluminescence intensity over this range was found to peak near 26\dgra and 34\dgra. At both of these angles, the 1550\,nm point was coincident with the band edge. Weak enhancement was noted at intermediate angles, which also coincided with points along the band edge. At 34\dgra, the intensity of the 550-nm peak was found to increase by approximately 26.6$\times$ in comparison to the weakest PL intensity across the range of measured angles. The enhancement was between 5$\times$ and 10$\times$ for the other \erbium peaks at this angle. The source of the disparity between the magnitude of enhancement for the various \erbium transitions noted in this experiment is unclear, but the result possibly indicates saturation of the transition efficiencies. The suggestion that the luminescence may be saturating at low efficiency---showing weak luminescence at high incident power---is evidence that substantial optimization of \erbium-doped PSi is necessary for it to be proposed as a relevant Si-based optoelectronic material.

Assessment of the relationship between the fine photonic features of the DBR and the luminescence variation observed in the experiment was hindered by the in-plane non-uniformity of the sample, which ``smear out'' the expected interference fringes. The results appear to be consistent with the angular absorption dependence seen in simulations, but a refined experiment is necessary to confirm this interpretation.

Despite the various challenges in fabrication and interpretation mentioned above, photonic-crystal-based UC enhancement in a in a spectral range relevant to c-Si photovoltaics has been demonstrated here for the first time. The method presented may serve as the basis for future experiments on slow-light enhancement in high-quality one-dimensional photonic crystals. 
%

%\bibliography{../../CMJBibliography/CraigJohnson}
%\bibliographystyle{apsrev4-1}
\end{document}